\documentclass[final,1p,times,twocolumn]{elsarticle}
\usepackage{graphicx,amssymb}
\journal{Physics Letters B}
\begin{document}
\begin{frontmatter}

\title{A test of unification towards the radio source PKS1413$+$135}
\author[inst1,inst2]{M. C. Ferreira}\ead{up200802537@fc.up.pt}
\author[inst1,inst3]{M. D. Juli\~ao}\ead{meinf12013@fe.up.pt}
\author[inst1]{C. J. A. P. Martins\corref{cor1}}\ead{Carlos.Martins@astro.up.pt}
\author[inst1,inst2,inst4]{A. M. R. V. L. Monteiro}\ead{mmonteiro@fc.up.pt}
\address[inst1]{Centro de Astrof\'{\i}sica, Universidade do Porto, Rua das Estrelas, 4150-762 Porto, Portugal}
\address[inst2]{Faculdade de Ci\^encias, Universidade do Porto, Rua do Campo Alegre, 4150-007 Porto, Portugal}
\address[inst3]{Faculdade de Engenharia, Universidade do Porto, Rua Dr Roberto Frias, 4200-465 Porto, Portugal}
\address[inst4]{Department of Applied Physics, Delft University of Technology, P.O. Box 5046, 2600 GA Delft, The Netherlands}
\cortext[cor1]{Corresponding author}

\begin{abstract}
We point out that existing astrophysical measurements of combinations of the fine-structure constant $\alpha$, the proton-to-electron mass ratio $\mu$ and the proton gyromagnetic ratio $g_p$ towards the radio source PKS1413$+$135 can be used to individually constrain each of these fundamental couplings. While the accuracy of the available measurements is not yet sufficient to test the spatial dipole scenario, our analysis serves as a proof of concept as new observational facilities will soon allow significantly more robust tests. Moreover, these measurements can also be used to obtain constraints on certain classes of unification scenarios, and we compare the constraints obtained for PKS1413$+$135 with those previously obtained from local atomic clock measurements.
\end{abstract}


\end{frontmatter}

\section{Introduction} 
\label{intro}

Nature is characterized by a set of physical laws and fundamental dimensionless couplings, and historically we have assumed that both of them are spacetime-invariant. For the former this is a cornerstone of the scientific method (it's hard to imagine how one could do science at all if it were not the case), but we must realize that for the latter the situation is different: there it is only a simplifying
assumption without further justification. It's remarkable how little we know about these couplings. We have no 'theory of constants', that describes their role in physical theories or even which of them are really fundamental. Indeed, our current working definition of a fundamental constant is simply: any parameter whose value cannot be calculated within a given theory, but must be found experimentally.

Fundamental couplings are known to \textit{run} with energy, and in many extensions of the standard model they will also \textit{roll} in time and \textit{ramble} in space (i.e., they will depend on the local environment). In particular, this will be the case in theories with additional spacetime dimensions, such as string theory. A detection of varying fundamental couplings will be revolutionary: it will automatically prove that the Einstein Equivalence Principle is violated (and therefore that gravity can't be purely geometry), and that there is a fifth force of nature.

Moreover, even improved null results are important. The simple way to understand this is to note that the natural scale for cosmological evolution of one of these couplings (if one assumes that it is driven by a fundamental scalar field) is the Hubble time. We would therefore expect a drift rate of the order of $10^{-10}$ yr${}^{-1}$. However, current local bounds coming from atomic clock comparison experiments \cite{Rosenband}, are already about 6 orders of magnitude stronger, and rule out otherwise viable dynamical dark energy models. 

Recent indications \cite{Dipole}, from quasar absorption systems observed with HIRES/Keck and UVES/VLT, suggest a parts-per-million spatial variation of the fine-structure constant $\alpha$ at low redshifts; although no known model can explain such a result without considerable fine-tuning, it should also be said that there is also no identified systematic effect that can explain it \cite{Thesis}.
One possible cause for concern is that almost all of the existing data has been taken with other purposes in mind (and subsequently reanalyzed for this purpose), whereas this kind of measurements need customized analysis pipelines and wavelength calibration procedures beyond those supplied by standard pipelines. An ongoing ESO UVES Large Program dedicated to fundamental physics will soon provide further measurements which may shed light on this issue. A review of the subject, including an overview of various astrophysical measurement techniques, can be found in \cite{Uzan}.

In the meantime, one can take the various existing measurements at face value, and try to ascertain whether they are consistent with one another. In this Letter we focus on three measurements of different combinations of the fine-structure constant $\alpha$, the proton-to-electron mass ratio $\mu$ and the proton gyromagnetic ratio $g_p$ towards the radio source PKS1413$+$135 at redshift $z\sim0.247$ \cite{Murphy,Darling,Kanekar2}. Together, these allow us to individually constrain each of these couplings. Although these constraints are relatively weak, our analysis serves two different purposes: apart from the aforementioned consistency check, it's also a proof of concept since forthcoming observational facilities will significantly improve existing measurements and allow for much stronger tests.

In addition to their intrinsic test as precision consistency tests of the standard cosmological model, these tests of the stability of fundamental constants can also be used to obtain constraints on certain classes of unification scenarios. This has been previously done for local (redshift $z=0$) tests, using comparisons of atomic clocks \cite{Clocks}. In this sense the current Letter is an extension of this formalism to the early universe. In a subsequent, longer paper we will extend this analysis to measurements at other redshifts.

\section{Varying couplings and unification}
\label{unify}

We wish to describe phenomenologically a class of models with simultaneous variations of several fundamental couplings, in particular the fine-structure constant $\alpha=e^2/\hbar c$, the proton-to-electron mass ratio $\mu=m_p/m_e$ and the proton gyromagnetic ratio $g_p$. The simplest way to do this is to relate the various changes to those of a particular dimensionless coupling, typically $\alpha$. Then if $\alpha=\alpha_0(1+\delta_\alpha)$ and
\begin{equation}
\frac{\Delta X}{X}=k_X\frac{\Delta\alpha}{\alpha}
\end{equation}
we have $X=X_0(1+k_X\delta_\alpha)$ and so forth.

The relations between the couplings will be model-dependent. We follow the analysis of \cite{Coc,Luo}, considering a class of grand unification models in which the weak scale is determined by dimensional transmutation and further assuming that relative variation of all the Yukawa couplings is the same. Finally we assume that the variation of the couplings is driven by a dilaton-type scalar field (as in \cite{Campbell}). For our purposes it's natural to assume that particle masses and the QCD scale vary, while the Planck mass is fixed. We then have
\begin{equation}
\frac{\Delta m_e}{m_e}=\frac{1}{2}(1+S)\frac{\Delta\alpha}{\alpha}
\end{equation}
(since the mass is simply the product of the Higgs VEV and the corresponding Yukawa coupling) and
\begin{equation}
\frac{\Delta m_p}{m_p}=[0.8R+0.2(1+S)]\frac{\Delta\alpha}{\alpha}\,.
\end{equation}
The latter equation is the more model-dependent one, as it requires modeling of the proton. At a phenomenological level, the choice $S=-1$, $R=0$ can also describe the limiting case where $\alpha$ varies but the masses don't.

With these assumptions one obtains that the variations of $\mu$ and $\alpha$ are related through
\begin{equation}\label{mualpha}
\frac{\Delta\mu}{\mu}=[0.8R-0.3(1+S)]\frac{\Delta\alpha}{\alpha}\,,
\end{equation}
where $R$ and $S$ can be taken as free phenomenological (model-dependent) parameters. Their absolute value can be anything from order unity to several hundreds, although physically one usually expects them to be positive. (Nevertheless, for our present purposes they can be taken as free parameters to be constrained by data.) Further useful relations can be obtained \cite{flambaum1,flambaum2,flambaum3} for the proton g-factor,
\begin{equation}\label{galpha}
\frac{\Delta g_p}{g_p}=[0.10R-0.04(1+S)]\frac{\Delta\alpha}{\alpha}\,.
\end{equation}
Together, these allow us to transform any measurement of a combination of constants into a constraint on the $(\alpha,R,S)$ parameter space.

\section{Measurements towards PKS1413$+$135}
\label{qsodata}

The spectrum  of the radio source PKS1413$+$135 includes a number of interesting molecular absorption as well as emission lines; the source is at the center of an edge-on spiral galaxy, and the absorption occurs in the galaxy's disk. From comparisons of different lines one can obtain measurements of several combinations of the fundamental couplings $\alpha$, $\mu$ and $g_p$. The underlying analysis methods are beyond the scope of this Letter, but the reader is referred to \cite{Uzan} for an overview of the subject. Specifically, three different (independent) measurements exist, which are summarized in Table \ref{table1}:
\begin{itemize}
\item Murphy {\it et al.} \cite{Murphy} use a comparison of 21cm HI absorption with molecular rotation absorption lines, which is sensitive to ${\alpha^{2}g_{p}}$
\item Darling \cite{Darling} uses a combination of 18cm OH lines and 21cm HI lines which is sensitive to ${\alpha^{2\times1.57}g_{p}\mu^{1.57}}$, cf. \cite{Darling0}. (Note that the constraint on $\alpha$ alone discussed in the paper is obtained if one assumes that only $\alpha$ varies while $\mu$ and $g_p$ are constant.)
\item Kanekar {\it et al.} \cite{Kanekar2} use conjugate 18cm OH lines which are sensitive to ${\alpha^{2\times1.85}g_{p}\mu^{1.85}}$ cf. \cite{Chengalur}.
\end{itemize}
Note that the first two are null results while the last measurement is a detection at more than two standard deviations. In all cases we define relative variations as
\begin{equation}
\frac{\Delta Q}{Q}=\frac{Q(z=0.247)-Q(z=0)}{Q(z=0)}\,.
\end{equation}

\begin{table}
\begin{center}
\begin{tabular}{|c|c|c|}
\hline
 $Q_{AB}$  & ${ \Delta Q_{AB}}/{Q_{AB}}$ & Reference \\ 
 \hline
 \hline
 ${\alpha^{2}g_{p}}$  & $(-2.0\pm4.4)\times10^{-6}$ & Murphy {\it et al.} \protect\cite{Murphy} \\
\hline
 ${\alpha^{2\times1.57}g_{p}\mu^{1.57}}$  & $(5.1\pm12.6)\times10^{-6}$ & Darling \protect\cite{Darling} \\
\hline
 ${\alpha^{2\times1.85}g_{p}\mu^{1.85}}$  & $(-11.8\pm4.6)\times10^{-6}$ & Kanekar {\it et al.} \protect\cite{Kanekar2} \\
\hline
\end{tabular}
\caption{\label{table1} Current combined measurements (with one-sigma uncertainties) at $z\sim0.247$ towards the radio source PKS1413$+$135.}
\end{center}
\end{table}

\begin{figure}
\begin{center}
\includegraphics[width=4in]{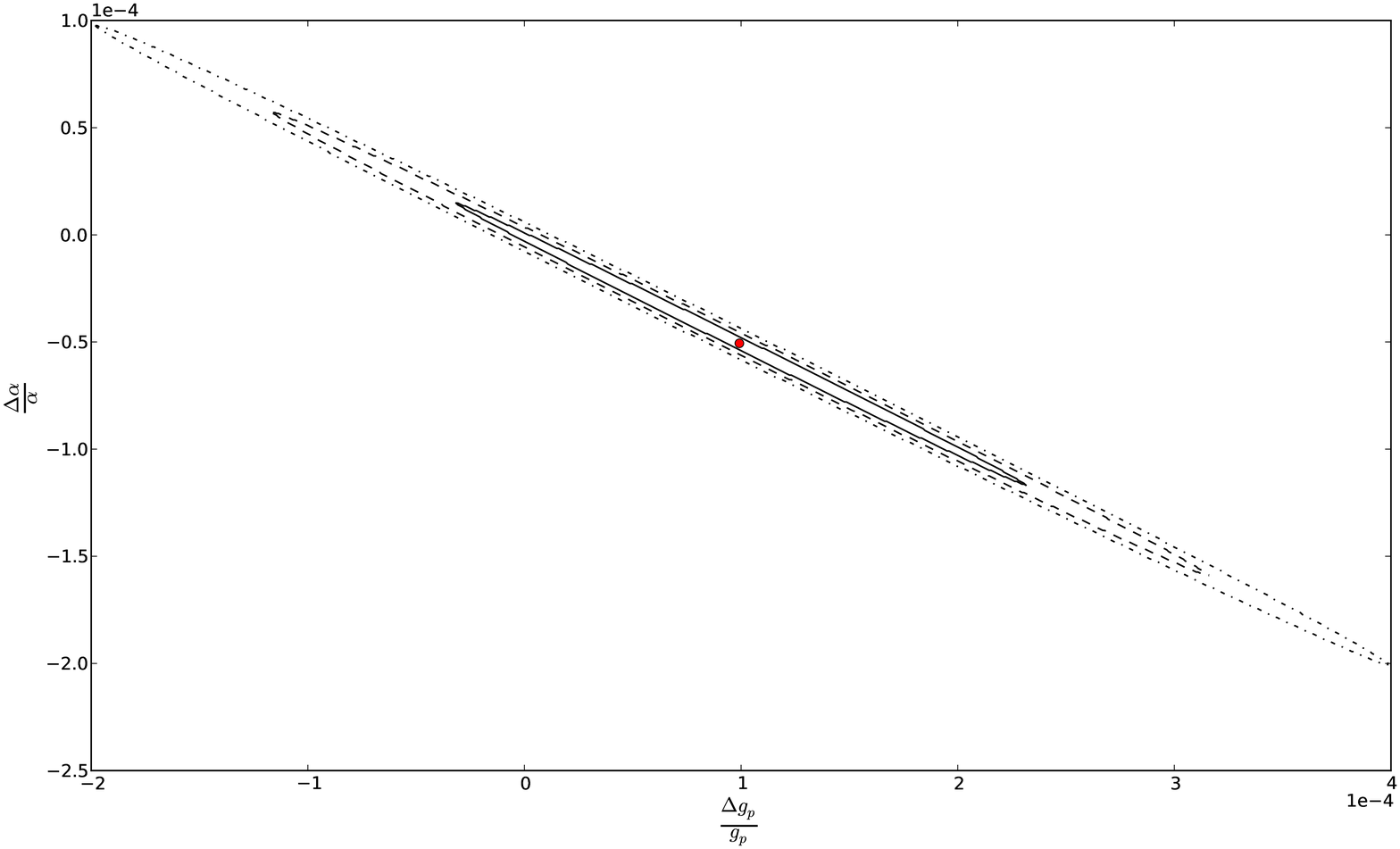}
\includegraphics[width=4in]{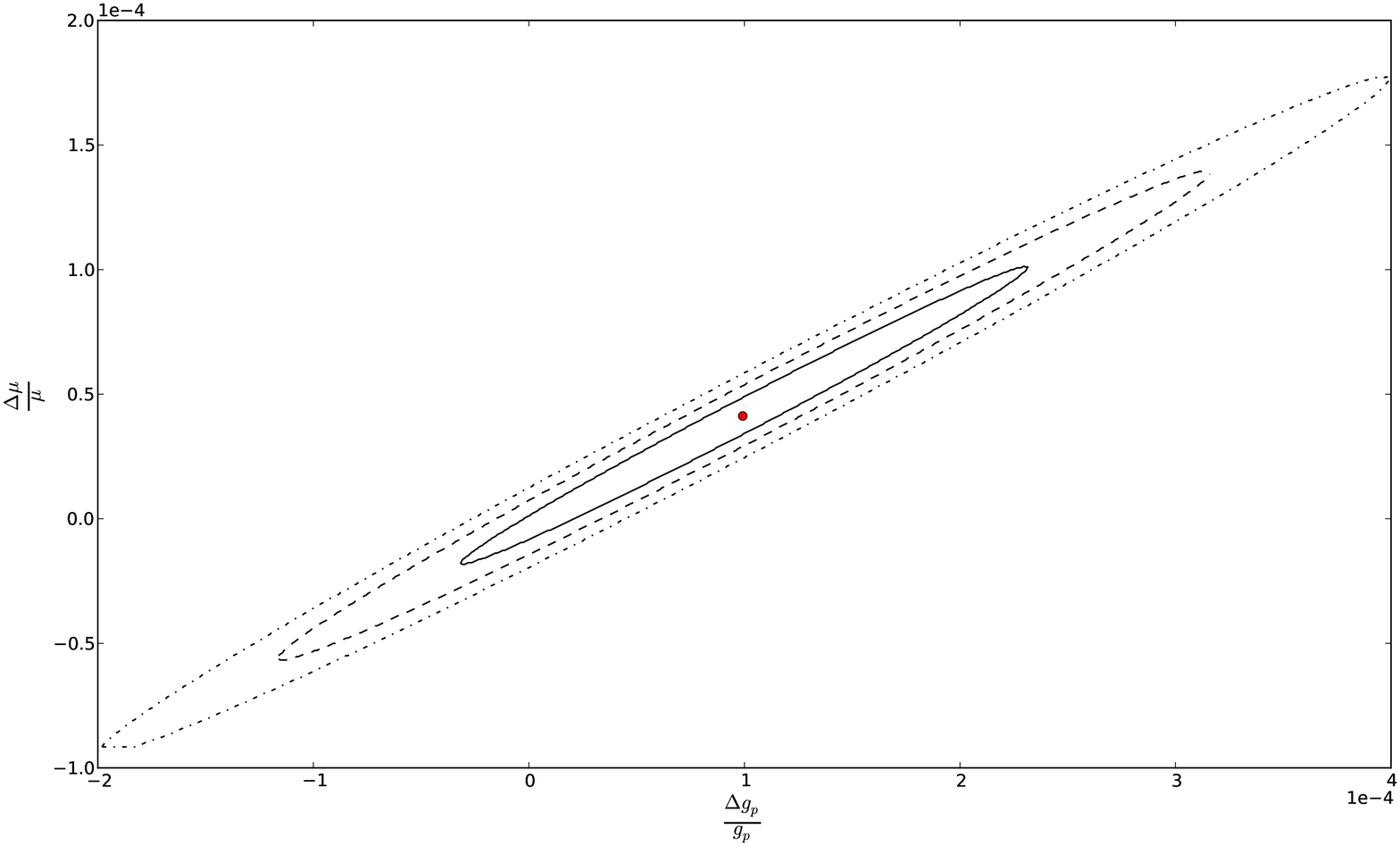}
\includegraphics[width=4in]{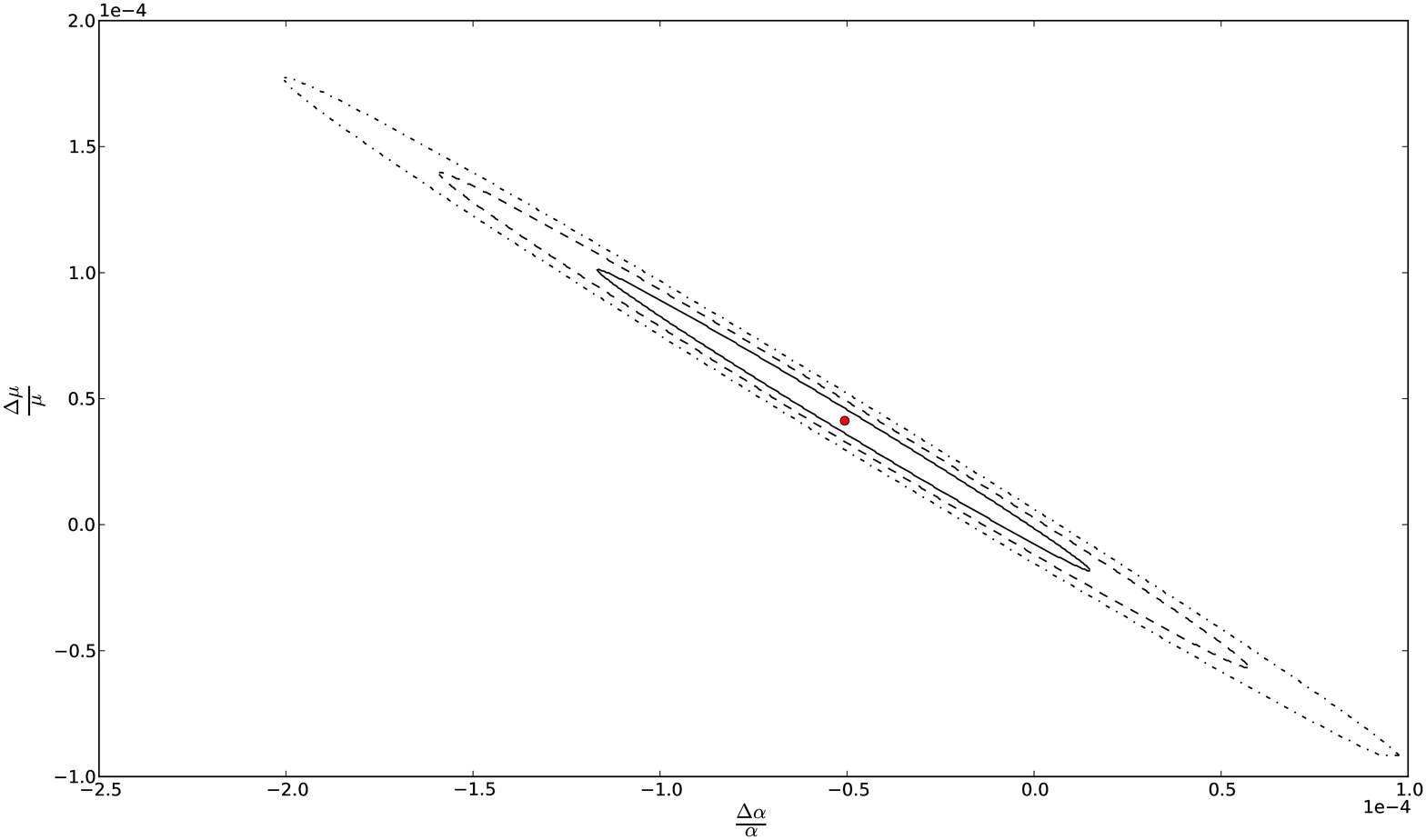}
\end{center}
\caption{\label{fig1}Two-dimensional likelihood contours for the relative variations of $\alpha$, $\mu$ and $g_p$ between $z=0.247$ and the local value ($z=0$). Solid, dashed and dotted lines correspond to one- , two- and three-sigma contours ($68.3\%$, $95.4\%$ and $99.97\%$ likelihood, respectively).}
\end{figure}

\begin{figure}
\begin{center}
\includegraphics[width=5in,angle=0]{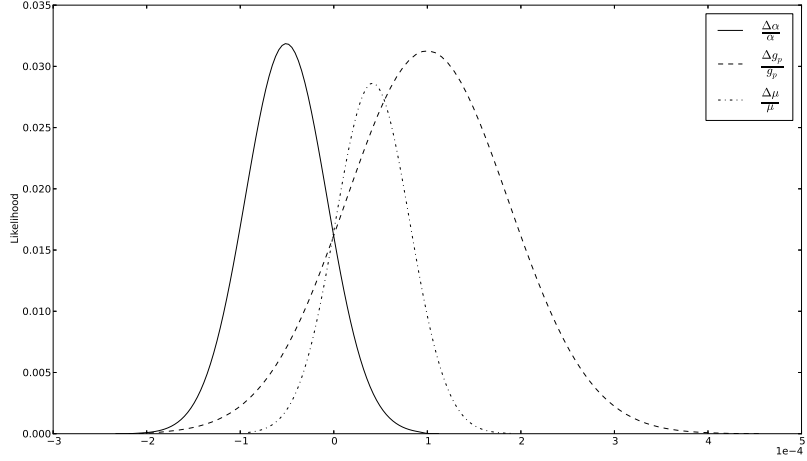}
\end{center}
\caption{\label{fig2}One-dimensional relative likelihoods (marginalized over the other quantities) for the relative variations of $\alpha$, $\mu$ and $g_p$ between $z=0.247$ and the local value ($z=0$).}
\end{figure}

From the three measurements we can obtain individual bounds on the variation of each of the couplings. Figure \ref{fig1} shows the two-dimensional likelihood contours in all relevant planes; notice the obvious degeneracies between the three parameters. The corresponding one-dimensional relative likelihoods are shown in figure \ref{fig1}. At the one-sigma ($68.3\%$) confidence level we find
\begin{equation}\label{result1}
\frac{\Delta\alpha}{\alpha}=(-5.1\pm4.3)\times10^{-5}\,
\end{equation}
\begin{equation}\label{result2}
\frac{\Delta\mu}{\mu}=(4.1\pm3.9)\times10^{-5}\,
\end{equation}
\begin{equation}\label{result3}
\frac{\Delta g_p}{g_p}=(9.9\pm8.6)\times10^{-5}\,,
\end{equation}
and at the two-sigma level all are consistent with a null result. These constraints are still relatively weak, and in particular do not yet provide a test of the dipole of Webb {\it et al.} \cite{Dipole}. However, improvements by one order of magnitude in each of the combined measurements (which are well within the reach of forthcoming facilities) should turn this into a stringent test. 

\section{Constraints on unification}
\label{unif}

\begin{figure}
\begin{center}
\includegraphics[width=5in]{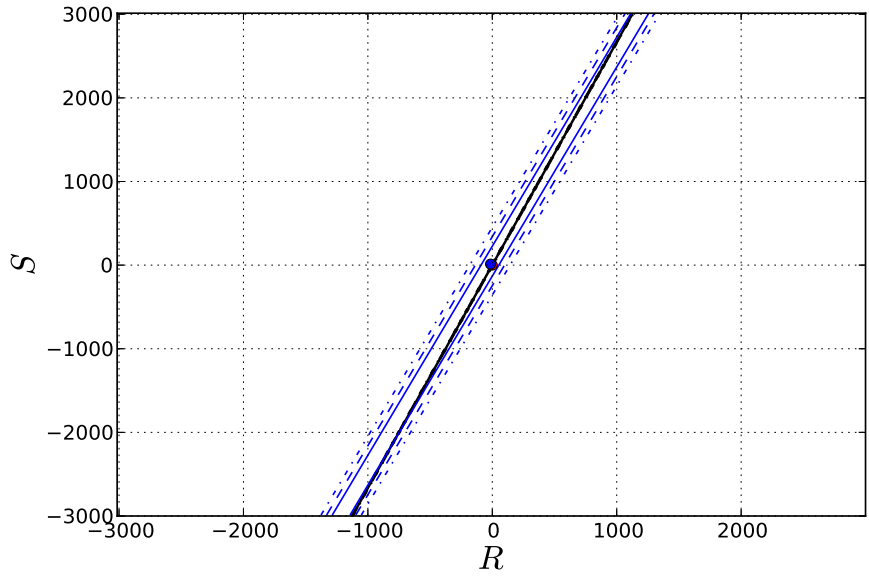}
\includegraphics[width=5in]{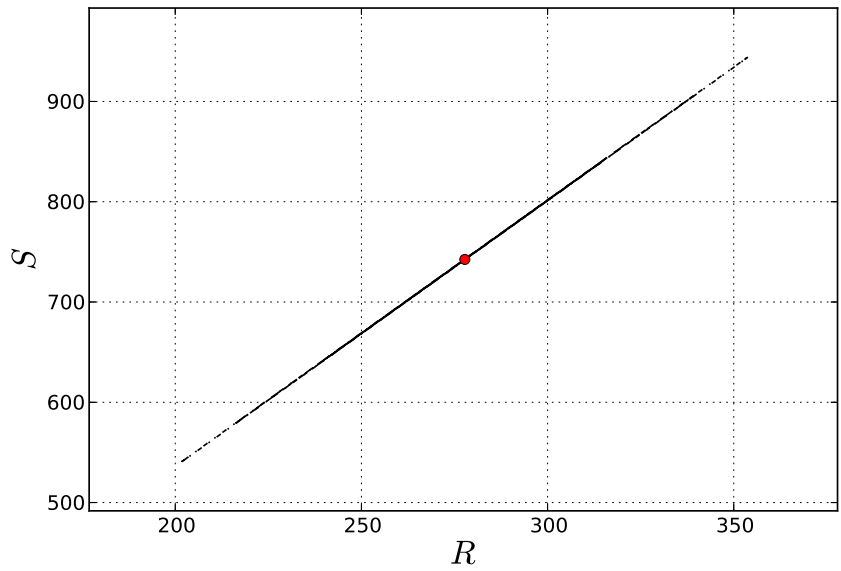}
\end{center}
\caption{\label{fig3}Two-dimensional likelihood contours in the $R$-$S$ plane; the top panel separately shows the degeneracy directions singled out by the $\mu$-$\alpha$ and $g_p$-$\alpha$ relations (the latter one having larger error bars), while the bottom one shows the combined contours. Solid, dashed and dotted lines correspond to one- , two- and three-sigma contours ($68.3\%$, $95.4\%$ and $99.97\%$ likelihood, respectively).}
\end{figure}

\begin{figure}
\begin{center}
\includegraphics[width=5in]{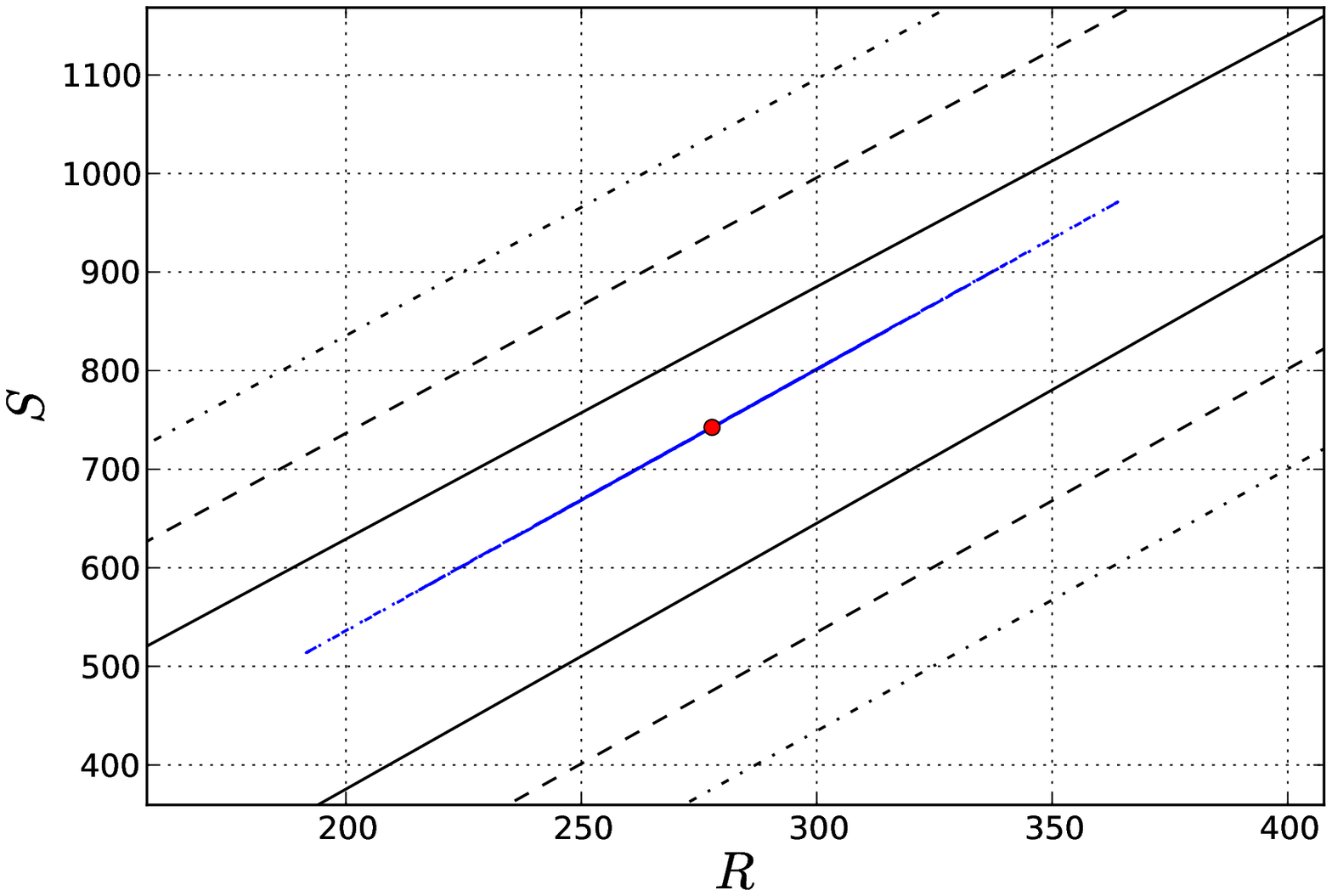}
\end{center}
\caption{\label{fig4}Two-dimensional likelihood contours in the $R$-$S$ plane. The broader contours correspond to constraints at $z=0$, coming from atomic clocks and discussed in \protect\cite{Clocks}. The smaller contours correspond to the combination of the atomic clock constraints with the ones for PKS1413$+$135. Solid, dashed and dotted lines correspond to one- , two- and three-sigma contours ($68.3\%$, $95.4\%$ and $99.97\%$ likelihood, respectively).}
\end{figure}

The bounds obtained in the previous section can now be translated, using Eqns. (\ref{mualpha}-\ref{galpha}) into constraints on the phenomenological unification parameters $R$ and $S$. Indeed, the astrophysical measurements restrict the two functions of $R$ and $S$ to be
\begin{equation}\label{mualpha2}
0.8R-0.3(1+S)=-0.81\pm0.85\,,
\end{equation}
\begin{equation}\label{galpha2}
0.10R-0.04(1+S)=-1.96\pm1.79\,.
\end{equation}

Figure \ref{fig3} shows relevant contours in the $R$-$S$ plane, assuming the values of Eqs. (\ref{result1}-\ref{result3}). Each of the above relations will determine a degeneracy direction in this plane, which are displayed in the first of the figure's panels; the combination of the two is then shown in the bottom panel. It's easy to show (simply by solving the above equations) that the most likely values of $R$ and $S$ given the current data are approximately
\begin{equation}\label{minR}
R\sim277.8\,
\end{equation}
\begin{equation}\label{minS}
S\sim742.5\,.
\end{equation}
although as the figure shows a strong degeneracy is still present. Interestingly this degeneracy direction is effectively the same as the one obtained from atomic clock measurements at $z=0$ and obtained in \cite{Clocks}. Figure \ref{fig4} shows the likelihood contours for $R$-$S$ in this case, as well as the result of the combination of the atomic clocks and PKS1413$+$135. The latter results dominate the analysis, and effectively select a sub-region from the band in the $R$-$S$ plane defined by the atomic clock data.

From this analysis we can finally obtain the best-fit values for $R$-$S$; at the one-sigma confidence level we obtain
\begin{equation}\label{minRR}
R=277\pm24\,
\end{equation}
\begin{equation}\label{minSS}
S=742\pm65\,.
\end{equation}
Although the notion that there is a 'standard' model for unification is debatable, it has been argued---particularly in \cite{Coc} and references therein---that typical values for these parameters are $R\sim30$ and $S\sim160$ (although these values certainly include a degree of uncertainty). Current constraints from atomic clocks \cite{Clocks} are fully consistent with these values, but our present analysis shows that this is not the case for PKS1413$+$135. 

\section{Conclusions}
\label{concl}

We have used available measurements of the values several combinations of dimensionless fundamental couplings towards PKS1413$+$135 to obtain individual constraints of the variations of $\alpha$, $\mu$ and $g_p$. The precision of the available measurements is not yet sufficient to provide a useful test of the spatial dipole scenario \cite{Dipole}.
However, with a next generation of observational facilities becoming available, this source can eventually provide a useful test.

We have also used our results to derive constraints on the class of unification scenarios described in \cite{Coc}, and compared them with those previously obtained from local atomic clock measurements. Analogous tests were performed in \cite{ref17,ref18,ref19}, with some assumptions on the redshift evolution of the variations. Our analysis, although focused on a specific class of models, has the conceptual advantage that no such further assumptions are needed.

Our analysis shows that both types of measurements prefer unification models characterized by a particular combination of the phenomenological parameters $R$ and $S$, with the PKS1413$+$135 providing stronger constraints on these parameters. It is noteworthy that the parameter values preferred by the current data do not coincide with (arguably naive) expectations on unification scenarios. A discussion of the causes and implications of this result is beyond the scope of the present work.

In any case our results motivate the interest of further, more precise measurements of fundamental couplings towards this and other similar astrophysical sources. More generally, they also highlight the point that the early universe is an ideal laboratory in which to carry out precision consistency tests of our standard cosmological paradigm and search for and constrain new physics. Future facilities such as ALMA, the E-ELT, the SKA and others will play a key role in this endeavor.

\section*{Acknowledgments}
This work was done in the context of the project PTDC/FIS/111725/2009 from FCT (Portugal), with additional support from grant PP-IJUP2011-212 (funded by Universidade do Porto and Santander-Totta). 

\bibliographystyle{model1-num-names}
\bibliography{pks}

\end{document}